\documentclass[12pt]{article}
\usepackage{graphicx}

\newcommand{\gsim}{ \mathop{}_{\textstyle \sim}^{\textstyle >} }

\begin{document}

\begin{titlepage}

\begin{center}

\hfill DESY 06-033

\vskip 5mm

{\Large\bf Constraint on Right-Handed Squark Mixings \\
  from $B_s-\bar B_s$ Mass Difference}

\vskip 7.5mm

{Motoi Endo$^{1,2}$ and Satoshi Mishima$^3$}

\vskip 7.5mm

{\it $^1$Deutsches Elektronen Synchrotron DESY, Notkestrasse 85,
  22607 Hamburg, Germany\\
  $^2$Institute for Cosmic Ray Research,
  University of Tokyo, Chiba 277-8582, Japan\\
  $^3$School of Natural Sciences, Institute for Advanced Study,
  Princeton, NJ 08540, U.S.A.
  }

\end{center}

\vskip 1.5mm

\begin{abstract}
\noindent
We point out that the right-handed squark mixings can sizably enhance 
SUSY contributions to $\Delta M_s$ by taking into account renormalization 
group effects via the CKM matrix. 
The recent result of $\Delta M_s$ from the {D\O} experiment at the 
Tevatron thus implies a strong constraint on the right-handed mixings. 
\end{abstract}

\end{titlepage}


Effects of the underlying physics at high energy scale are imprinted not
only in flavor structures of the matters in the standard model (SM), but
also in those of their superpartners by extending the SM to include
supersymmetry (SUSY).  
They evolve from the cutoff scale to a low energy via the renormalization 
running, and are recognized through signals of the flavor-changing neutral 
currents (FCNCs). 

The $\tilde b - \tilde s$ mixings of the right-handed squarks 
have attracted a lot of interests~\cite{Moroi:2000tk} in the light 
of the discovery of the neutrino oscillations with large mixing 
angles~\cite{Fukuda:1998mi}, and the success of the supersymmetric 
grand unification. 
The mixings are parameterized by the flavor-changing components between 
$\tilde b_R-\tilde s_R$ and $\tilde b_L-\tilde s_R$ in the mass matrices, 
which are called the $RR$ and $RL$ mixings, respectively. At the weak scale, 
they contribute to $b \to s$ transition processes. Some golden modes 
have been already measured precisely in experiments, and their results 
are compared to the SM predictions. Particularly, the measured branching 
ratio of the inclusive $B_d \to X_s \gamma$ decay is known to agree well 
with the SM value~\cite{Group:2006bi,Asatrian:2005yr}. Thus it  
provides one of the severest constraints on the down-type squark mixings, 
including the $RR$ and $RL$ mixings. Even with the constraint from 
${\rm Br}(b \to s\gamma)$, 
there is still left large possibility to detect sizable effects on some 
$b \to s$ processes, especially, from the right-handed squark mixings. 

Recently, the {D\O} collaboration have reported the updated result of 
the mass difference of the $B_s$ mesons~\cite{Abazov:2006dm}:
\begin{eqnarray}
  17\,{\rm ps}^{-1} \;<\; \Delta M_s \;<\; 21\,{\rm ps}^{-1} 
  ~~~90\%\ {\rm C.L.}\ ,
  \label{eq:deltams_exp}
\end{eqnarray}
which is the first result with a direct two-side bound.
Although the data includes large uncertainties, this result 
is in agreement with the SM predictions, which are estimated 
as $21.3 \pm 2.6$ ps$^{-1}$ by the UTfit group~\cite{Bona:2005vz} 
and $20.9^{+4.5}_{-4.2}$ ps$^{-1}$ by the CKMfitter group~\cite{Charles:2004jd}.
In the supersymmetric SM, it is known that a combination of the $LL$ and 
$RR$ mixings, $(m_{\tilde d_{LL}}^2)_{23} (m_{\tilde d_{RR}}^2)_{23}$, can enhance 
the SUSY contributions to $\Delta M_s$ sizably~\cite{Ball:2003se}. 
If the $LL$ mixing is suppressed sufficiently, the current data of 
$\Delta M_s$ remains insensitive to the right-handed 
mixings~\cite{Ciuchini:2006dx}. 

Among the squark mixings, the $LL$ mixing at the weak scale generally 
receives a correction of at least $O(0.01)$. This is because the SUSY 
breaking effects are usually mediated to the visible sector at the high 
energy scale. Actually in general supersymmetric SM at the weak scale, 
we might choose the down-type $LL$ squark mixing to vanish. However, 
in realistic models, the mixing will be induced because the CKM matrix 
affects the left-handed mass matrix of the down-type squarks during the 
renormalization group evolutions. In a class of supergravity mediations, 
the $LL$ mixing generally gets a correction of $(\delta_{LL}^d)_{23} \sim 
\lambda^2$, where $\lambda \sim 0.2$ is the Wolfenstein parameter. 
In this letter, we want to emphasize that this tiny mixing is significant 
for $\Delta M_s$ when we discuss the mixings in the right-handed sector. 
We will show that even without any imprinted $LL$ mixing, the $RR$ squark 
mixing can affect $\Delta M_s$ to the level of the SM value satisfying 
with the bound from ${\rm Br}(b \to s \gamma)$. 

Let us first review the SUSY contributions to $\Delta M_s$. 
The $B_s-\bar B_s$ transition is represented by the transition matrix element,
\begin{eqnarray}
  M_{12} \;=\; M_{12}^{\rm SM} + M_{12}^{\rm SUSY} \;\equiv\; M_{12}^{\rm SM} (1 + R).
\end{eqnarray}
In terms of $R$, which corresponds to the SUSY contributions, 
the mass difference between $B_s$ and $\bar B_s$ becomes $\Delta M_s 
= \Delta M_s^{\rm SM} |1 + R|$. Although the estimation of the SM value 
contains large hadronic uncertainties, the ratio $\Delta M_s/\Delta M_d$ 
can be predicted more cleanly. Then $R$ is given by
\begin{eqnarray}
  |1 + R| \;=\; 
  \frac{\Delta M_d^{\rm SM}}{\Delta M_s^{\rm SM}}
  \frac{\Delta M_s^{\rm EXP}}{\Delta M_d^{\rm EXP}}
  \;=\;
  \frac{M_{B_d}}{M_{B_s}} \frac{1}{\xi^2} \left| \frac{V_{td}}{V_{ts}} \right|^2
  \frac{\Delta M_s^{\rm EXP}}{\Delta M_d^{\rm EXP}},
  \label{eq:deltams_deltamd}
\end{eqnarray}
where $M_{B_{d,s}}$ are masses of the mesons, $M_{B_d} = 5.279$ GeV and 
$M_{B_s} = 5.375$ GeV, and $\xi \equiv f_{B_s}/f_{B_d} \sqrt{B_{B_s}/B_{B_d}} 
= 1.24 \pm 0.04 \pm 0.06$ is defined by ratios of the decay constants 
and of the bag parameters~\cite{Battaglia:2003in}. 
In above expression, we assume $\Delta M_d^{\rm EXP} = \Delta M_d^{\rm SM}$ 
because the SUSY contributions to $b \to d$ transitions is tightly limited 
by the experimental results~\cite{Charles:2004jd}~\footnote{
  Although the ratio $\Delta M_d^{\rm SM}/\Delta M_d^{\rm EXP}$ may include 
  SUSY contributions at $\sim 10\%$~\cite{Charles:2004jd}, the following 
  discussions remain valid. Similarly, even when we use $\Delta M_s$ itself 
  instead of $\Delta M_s/\Delta M_d$, the results are almost same.
}. From the experimental result in Eq.~(\ref{eq:deltams_exp}) and 
$\Delta M_d^{\rm EXP} = 0.507 \pm 0.004$ ps$^{-1}$ \cite{Group:2006bi}, 
the SUSY contributions are favored to be in the region, 
\begin{eqnarray}
  0.55 \;<\; |1 + R| \;<\; 1.37\ ,
\end{eqnarray}
with including a $40$ \% uncertainty from the SM estimations, 
which is mainly due to the ratio of the CKM matrix elements 
$V_{td}/V_{ts}$~\cite{Bona:2005vz,Charles:2004jd}.

The SM contribution, $M_{12}^{\rm SM}$, is obtained by exchanging the $W$ boson 
and top quark. On the other hand, the SUSY contribution $M_{12}^{\rm SUSY}$ is  
by exchanging the gluino and down-type squarks. In the following analysis, 
we evaluate the SUSY contributions by the full expressions, namely, without 
the mass-insertion approximation~\cite{Hall:1985dx,Gabbiani:1996hi} 
(see Ref.~\cite{Harnik:2002vs} for the explicit forms which are relevant 
for the right-handed mixings). In contrast, it is known that the chargino 
contributions are much suppressed compared to the gluino 
ones~\cite{Ball:2003se}. Thus we neglect them in the following. 

The flavor violations are caused by the intermediating squarks 
in the diagrams. The mass matrix of the down-type squarks contains 
the flavor-changing components with/without chirality flipping, 
and the terms relevant for the $b \to s$ transitions are 
$(m_{\tilde d_{RR(LL)}}^2)_{23}$ at the mixing between $\tilde s_{R(L)}^*$ 
and $\tilde b_{R(L)}$, and $(m_{\tilde d_{RL(LR)}}^2)_{23}$ between $\tilde s_{R(L)}^*$ 
and $\tilde b_{L(R)}$, where $L,R$ correspond to the chirality of the squarks, 
and the numbers outside the parenthesis means the mixing components between 
the second and third generations. The $B_s-\bar B_s$ transition is then induced 
by pairs of the mixings: the $RR(LL)$ and $RL(LR)$ mixings, respectively. 
It is important to mention that the ratio $R$ is enhanced especially by 
a pair of the $LL$ and $RR$ mixings, $(m_{\tilde d_{LL}}^2)_{23} \times 
(m_{\tilde d_{RR}}^2)_{23}$~\cite{Ball:2003se}. 

The $LL$ squark mixing receives the radiative corrections via the CKM matrix 
during the renormalization group evolution. The running from the cutoff scale 
$M_X$ to the weak scale $M_W$ gives a mixing such as
\begin{eqnarray}
  (\delta_{LL}^d)_{23} \;\equiv\; \frac{(m_{\tilde d_{LL}}^2)_{23}}{m_{\tilde q}^2}
  \;\simeq\; -\frac{1}{8\pi^2} Y_t^2 V_{ts} 
  \frac{3m_0^2 + a_0^2}{m_0^2} \ln \frac{M_X}{M_W},
\end{eqnarray}
where $Y_t$ is the top Yukawa coupling, $m_0 \sim m_{\tilde q}$ are typical 
values of diagonal components of the scalar and squark mass matrices, and 
$a_0$ is a trilinear coupling, $A_t \equiv a_0 Y_t$. We stress that the 
mixing arises even when the mass matrix is diagonal at the cutoff scale. 
Actually, in supergravity mediations with $M_X$ at the Planck scale 
or the GUT scale $M_G \simeq 2 \times 10^{16}$ GeV, the renormalization 
group evolution induces $(\delta_{LL}^d)_{23} \simeq 0.04$. With this $LL$ 
mixing, the $RR$ squark mixing can contribute effectively to the ratio $R$. 
It should be stressed that the $LL$ mixing of $O(0.01)$ is rather general 
independent of the details of the squark mass matrices at the cutoff scale, 
and it is hard to suppress the $LL$ mixing at the weak scale unless the 
mixing is tuned at the cutoff scale. In the following analysis, we will
use $(\delta_{LL}^d)_{23} \simeq 0.04$ as a representative value~\footnote{
  If the $LL$ mixing of $\gsim O(0.01)$ is imprinted at the cutoff scale, 
  $M_{12}^{\rm SUSY}$ will be enhanced too much due to the interference with 
  the $RR$ mixing. The analysis, however, depends on the structure of 
  the mass matrices especially of the left-handed sector. In contrast, 
  the $LL$ mixing of $O(0.01)$ at the weak scale is independent of the 
  details of the models, and the following constraint on the right-handed 
  squark mixing is rather generic. 
}.

Let us consider the bound of the squark mixings from ${\rm Br}(b \to s\gamma)$.
The SUSY contributions to this mode are induced through the gluino, chargino 
and charged-Higgs loops~\footnote{
  Those diagrams also contribute to the decay amplitude of $b \to s ll$, 
  which is sensitive to the sign of $C_{7\gamma}$, and has been already 
  limited by the experiment~\cite{Ishikawa:2006fh}. The bound is always 
  satisfied here because the $LL$ mixing is small enough, 
  $(\delta_{LL}^d)_{23} = 0.04$~\cite{Lunghi:1999uk}. 
}. They significantly depend on $\tan\beta$ and a sign 
of the higgsino mass parameter, $\mu_H$. In fact, the SUSY contributions are 
enhanced by large $\tan\beta$, and sign$(\mu_H)$ determines signs of the 
chargino and gluino contributions compared to the SM and charged-Higgs ones. 
As well as the evaluations of $\Delta M_s$, we evaluate the SUSY contributions 
including the gluino ones by the full expressions~\cite{Ciuchini:1997xe,Degrassi:2000qf,Borzumati:2003rr}~\footnote{
  The single mass-insertion approximation of the gluino-mediated diagrams 
  is not consistent with the full estimations for the dipole operators, 
  $C_{7\gamma}$ and $C_{8G}$, when we study the $LL$ and/or $RR$ squark mixings. 
  Rather the dominant contributions are provided by so-called the double 
  mass-insertion diagrams even for small $\tan\beta$, like $\tan\beta = 5$. 
  See Ref.~\cite{Endo:2004dc} for the explicit forms of the double-mass 
  insertions. 
}. In the following analysis, we take the bound,
\begin{eqnarray}
  2.0 \times 10^{-4} \;<\; 
  {\rm Br}(b \to s\gamma)
  \;<\; 4.5 \times 10^{-4},
\end{eqnarray}
which is rather conservative after taking into account both the experimental 
and theoretical uncertainties. 

\begin{figure}[ht]
  \begin{center}
    \includegraphics{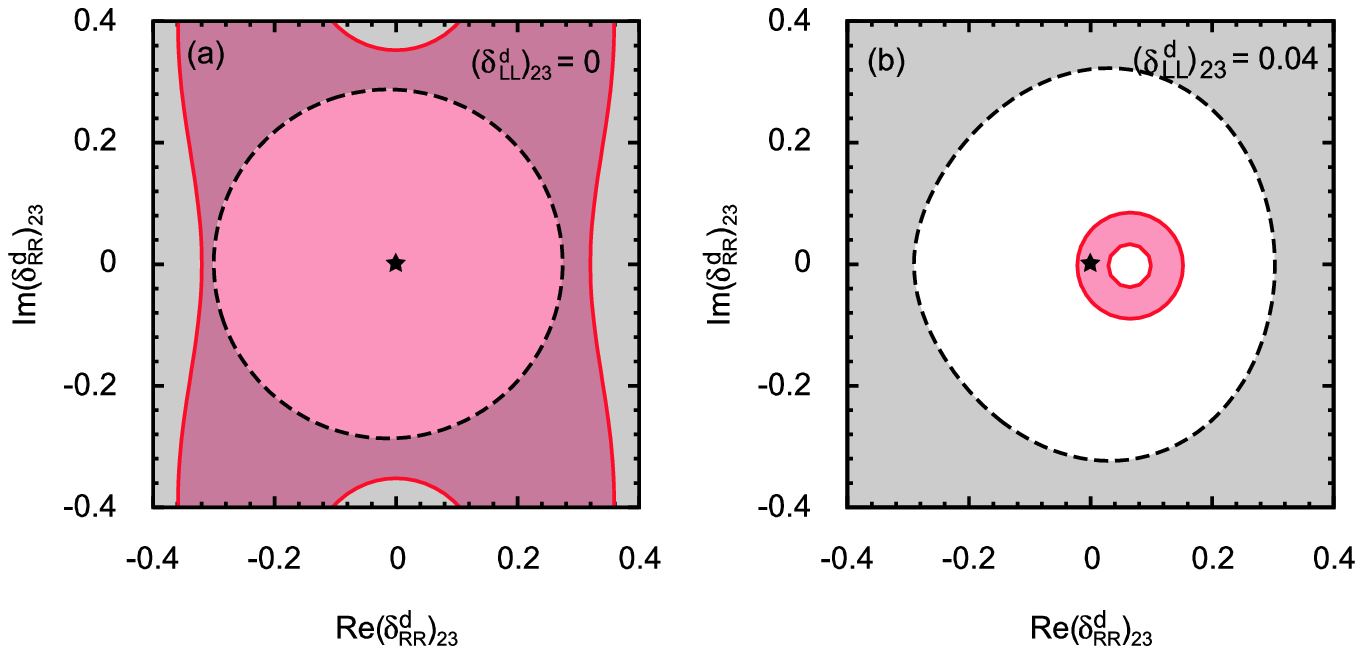}
  \end{center}
  \caption{
    Regions which reproduce the current experimental value of $\Delta M_s$ 
    for $(\delta_{LL}^d)_{23} = 0$ (a) and $(\delta_{LL}^d)_{23} = 0.04$ (b),
    with the uncertainties of 20\% due to the SM estimations (Red, enclosed 
    by solid lines). The star point corresponds to $(\delta_{RR}^d)_{23} = 0$. 
    The shadowed region outside the dashed line is excluded by ${\rm Br}(b 
    \to s \gamma)$. We take $m_{\tilde g} = m_{\tilde q} = 500$ GeV, 
    $\mu_H = 500$ GeV, and $\tan\beta = 10$. }
  \label{fig:DeltaMs_RR}
\end{figure}

We calculated the SUSY contributions to $\Delta M_s$ as well as ${\rm Br}(b 
\to s\gamma)$. In Fig.~\ref{fig:DeltaMs_RR}, we displayed the regions which 
is favored by the current result of $\Delta M_s$, and that is excluded by 
${\rm Br}(b \to s\gamma)$ for a range of the real and imaginary parts of 
the $RR$ mixing, $(\delta_{RR}^d)_{23} \equiv (m_{\tilde d_{RR}}^2)_{23}/m_{\tilde q}^2$, 
where $m_{\tilde q}^2$ is a typical mass of the squarks. Here we assume all 
relevant soft parameters including $\mu_H$ are $m_{\rm soft} = 500$ GeV, and 
$\tan\beta = 10$. In order to clarify the effects of the renormalization 
group evolutions, we show the result of $(\delta_{LL}^d)_{23} = 0$ in 
Fig.~\ref{fig:DeltaMs_RR}(a), and that of $(\delta_{LL}^d)_{23} = 0.04$ in 
Fig.~\ref{fig:DeltaMs_RR}(b). Consequently, we find that the radiative 
corrections in the $LL$ mixing can enhance the SUSY contributions for 
$\Delta M_s$ extremely to the extent of the magnitude which is implied 
by the current data (Fig.~\ref{fig:DeltaMs_RR}(b)), compared to the 
result without the effects (Fig.~\ref{fig:DeltaMs_RR}(a)). 

\begin{figure}[ht]
  \begin{center}
    \includegraphics{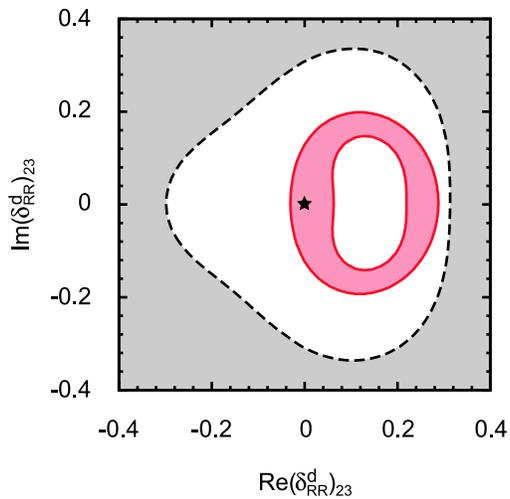}
  \end{center}
  \caption{
    Same as Fig.~\ref{fig:DeltaMs_RR}(b), but $m_{\tilde g} = 300$ GeV 
    and $m_{\tilde q} = 1$ TeV. }
  \label{fig:DeltaMs_HSq}
\end{figure}

We also considered other two sets of the mass spectrum of the gluino and 
squarks. The first pattern is $m_{\tilde g} \ll m_{\tilde q}$. This type is 
interesting for the FCNCs whose amplitude is dominated by the Wilson 
coefficient of the gluon-dipole operator, $O_{8G}$. Since the relevant 
contribution to ${\rm Br}(b \to s\gamma)$ comes from the photo-dipole 
operator, $O_{7\gamma}$, the mass pattern is favored to enhance the SUSY 
contributions for such FCNCs with satisfying the bound from ${\rm Br}
(b \to s\gamma)$, namely, enhance $C_{8G}$ compared to 
$C_{7\gamma}$~\cite{Harnik:2002vs}. We estimated numerically 
$\Delta M_s$ and ${\rm Br}(b \to s\gamma)$ in this case: $m_{\tilde g} \ll 
m_{\tilde q}$. In Fig.~\ref{fig:DeltaMs_HSq}, the parameters are set as the 
same as Fig.~\ref{fig:DeltaMs_RR}(b), but the gluino and squark masses are 
$m_{\tilde g} = 300$ GeV and $m_{\tilde q} = 1$ TeV, respectively. 
We find that although 
both the contributions are suppressed by the heavy squarks, the $\Delta M_s$ 
region remains inside that excluded by ${\rm Br}(b \to s\gamma)$. 

\begin{figure}[ht]
  \begin{center}
    \includegraphics{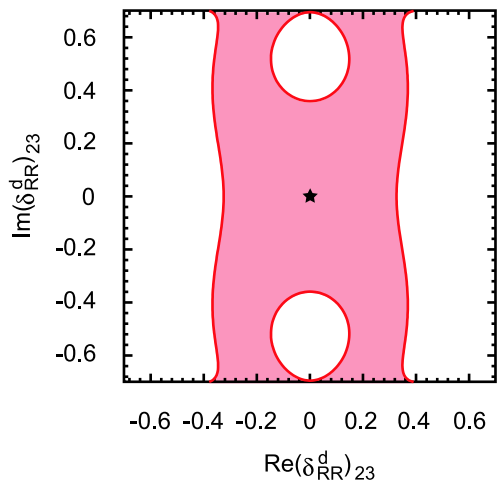}
  \end{center}
  \caption{
    Same as Fig.~\ref{fig:DeltaMs_RR}(b), but $m_{\tilde q_L}$ decoupled. }
  \label{fig:DeltaMs_HL}
\end{figure}

The second mass spectrum is $m_{\tilde q_L} \gg m_{\tilde q_R}$. Such a pattern 
can suppress the SUSY contributions to the electric dipole moments (EDMs). 
It has been pointed out that the strong bounds are imposed on the CP-violating 
$\tilde b-\tilde s$ mixings from the hadronic EDMs~\cite{Hisano:2003iw,Endo:2003te}.
Especially it is very strong for the right-handed sector even when 
we allow large hadronic uncertainties. In the above discussions, 
we implicitly assumed accidental cancellations among the additional 
SUSY contributions for the EDMs. Another way out of the EDMs is to 
suppress them by large squark masses. When the left-handed squarks decouple, 
the EDM bound can be relaxed to the negligible level. The numerical result 
is in Fig.~\ref{fig:DeltaMs_HL}, where the parameters are the same as 
Fig.~\ref{fig:DeltaMs_RR}(b) but $m_{\tilde q_L} \gg m_{\rm soft}$. We find 
the favored regions by $\Delta M_s$ in the graph, though all the constraints 
are satisfied because the SUSY contributions to ${\rm Br}(b \to s\gamma)$ 
as well as the EDMs can be neglected by large $m_{\tilde q_L}$. 

Even when we take other sets of parameters, the results remain similar. 
Let us first flip the sign of $\mu_H$ parameter. Then the cancellation 
becomes worse among the SUSY contributions to ${\rm Br}(b \to s\gamma)$. 
In contrast, $\Delta M_s$ is insensitive to the sign$(\mu_H)$. 
Then the favored region from $\Delta M_s$ starts to be restricted by 
${\rm Br}(b \to s\gamma)$. Instead, when we take $\tan\beta = 40$, 
the remaining contributions after the cancellation become enhanced for 
${\rm Br}(b \to s\gamma)$, while $\Delta M_s$ is insensitive to $\tan\beta$. 
We checked that the favored regions by the observables become narrower 
for the $RR$ mixing in both cases. 

There might be a large mixing in the $RL$ component at the cutoff scale. 
As already known, the flavor mixings with chirality flipping cannot induce 
large $\Delta M_s$ at the weak scale~\cite{Ball:2003se}. This is because 
the mixings are constrained by ${\rm Br}(b \to s\gamma)$ to the extent 
of the negligible level for $\Delta M_s$. On the contrary, the $RL$ mixing 
at the high energy scale can contribute sizably to the $RR$ mixing at the 
weak scale through the renormalization group running~\cite{Kane:2004ku}. 
Thus the similar arguments may be applicable for the $RL$ mixing at the 
cutoff scale as those of the $RR$ case, though the detailed discussion 
depends on the parameters. 
Consequently, we conclude that the SUSY contributions to $\Delta M_s$ is 
sensitive to the right-handed squark mixings at the cutoff scale in the
light of the current result of $\Delta M_s$ by considering the radiative
correction for the $LL$ squark mixing. 

Let us comment on implications to other $b \to s$ transition processes. 
The mixing-induced CP asymmetry of $b \to q\bar qs$ decays has been 
measured by BaBar and Belle~\cite{Group:2006bi}. Although the results 
still contain large theoretical/experimental uncertainties, we obtain an 
important implication from the measurements that all the values tend to 
shift to the same side from the SM values, which is determined by  
$b \to c\bar c s$ modes. This feature is observed independently of the 
parity of the final states. If the displacements are due to SUSY effects, 
the results naturally imply additional CP-violating mixings in the left-handed 
squark sector~\cite{Endo:2004dc}. On the contrary, in order to realize the 
same feature by the right-handed squark mixings, rather large SUSY 
contributions are required~\cite{Larson:2004ha}. 
We checked that such mixings induce too large $\Delta M_s$ to stay 
within the current data as long as we consider the renormalization 
group effects. This means that although the SUSY $SU(5)$ GUT + $\nu_R$ 
model is one of the best candidates which naturally induce large SUSY 
contributions to $b \to q\bar qs$ decays~\cite{Moroi:2000tk}, the model 
is disfavored in order to explain the current experimental result 
of $\Delta M_s$ as well as those modes. Other interesting $b \to s$ 
observables are mixing-induced CP asymmetry of the $b \to s\gamma$ 
decay and $B_s \to J/\psi \phi$. Since they are sensitive to the 
right-handed squark mixings, we can still expect to detect signals 
of new physics at LHC/super $B$-factory. We thus stress that
measurements of $\Delta M_s$ have impacts on the squark-flavor mixings. 

{\it Note Added}: shortly after this letter, the CDF collaboration 
reported the new result of $\Delta M_s$. The experimental error was 
reduced very well, and the result is found to be consistent with the 
{D\O} value, thus the SM estimation. Even taking into account the 
CDF result, the analysis in this letter does not change because the 
uncertainties in the analysis dominantly come from the SM estimation. 
Although the uncertainty may be reduced by combining the other 
measurements of the flavor changing processes, such an analysis 
is beyond the scope of this letter and should be studied elsewhere. 

\section*{Acknowledgements}

The authors would like to thank Ahmed Ali and Masahiro Yamaguchi 
for useful discussions and comments. 
M.E. would like to thank the Japan Society for Promotion of 
Science for financial support. 
This work was partly supported by the U.S. Department of Energy 
under Grant No. DE-FG02-90ER40542.

\clearpage

\newpage

\end{document}